\documentclass[prl,twocolumn,floatfix,showpacs ]{revtex4-1}
\UseRawInputEncoding
\usepackage{amsmath}
\usepackage{amssymb}
\usepackage{graphicx}% Include figure files
\usepackage{dcolumn}% Align table columns on the decimal point
\usepackage{bm}% bold math
\usepackage[colorlinks=true,linkcolor=blue,anchorcolor=blue, citecolor=cyan,urlcolor=cyan]{hyperref}% add hypertext capabilities
\usepackage[mathlines]{lineno}% Enable numbering of text and display math
\usepackage{ulem}
\usepackage{epstopdf}
\usepackage{tabularx} % ¼ÓÔØ tabularx ºê°ü
\begin{document}
%========================================================
\title{Distinguishing apparent and hidden altermagnetism via uniaxial strain in $\mathrm{CsV_2Te_2O}$-family}%   in symmetry-enforced net-zero-magnetization magnets}
\author{San-Dong Guo}
\email{sandongyuwang@163.com}
\affiliation{School of Electronic Engineering, Xi'an University of Posts and Telecommunications, Xi'an 710121, China}
\author{Yang Liu}
%\email{yangliuphys@zju.edu.cn}
\affiliation{School of Physics, Zhejiang University, Hangzhou 310058, China}
%\affiliation{$^3$Key Laboratory of Extraordinary Bond Engineering and Advanced Materials Technology of Chongqing, School of Electronic Information Engineering, Yangtze Normal University, Chongqing 408100, China}
%\affiliation{ address 3}
%\date{\today}
%=======================Abstract===============================================================
\begin{abstract}
The hidden altermagnetism has been theoretically proposed and then experimentally confirmed in metal $\mathrm{Cs_{1-\delta}V_2Te_2O}$, which  exhibits two nearly degenerate ground-state magnetic configurations (C-type and G-type) corresponding respectively to apparent and hidden altermagnetism.
Here, we propose that in-plane uniaxial strain can be utilized to distinguish apparent and hidden altermagnetism. Under uniaxial strain, apparent altermagnetism exhibits an obvious net magnetic moment, whereas hidden altermagnetism maintains zero net magnetic moment. The magnetic moment induced by uniaxial strain here, namely the piezomagnetic effect, differs from that in semiconductors, where strain must be applied first followed by carrier doping to generate net magnetism.
First-principles calculations verify our proposal, revealing that the magnetic moment induced by uniaxial strain in C-type antiferromagnetic  $\mathrm{CsV_2Te_2O}$ is much larger than that in the previously studied altermagnetic semiconductors.
Furthermore, we also investigate the electronic state transitions of semiconductors  featuring a crystal structure analogous to $\mathrm{CsV_2Te_2O}$  under uniaxial strain, and verify our proposal in specific material via first-principles calculations.
Our work provides an experimentally feasible strategy to distinguish apparent and hidden altermagnetism in material $\mathrm{Cs_{1-\delta}V_2Te_2O}$, and extends the physical implication of the piezomagnetic effect, which can be directly verified in experimentally synthesizable  $\mathrm{KV_2Se_2O}$ and  $\mathrm{Rb_{1-\delta}V_2Te_2O}$.

\end{abstract}
\maketitle
%\tableofcontents
%========================Introduction===========================================================
\textcolor[rgb]{0.00,0.00,1.00}{\textbf{Introduction.---}}
Altermagnetism combines properties of both ferromagnets and antiferromagnets\cite{k4,k5}. Like antiferromagnets, altermagnets have zero net magnetization due to compensated magnetic sublattices. However, they exhibit spin-splitting electronic bands-similar to ferromagnets-where spin-up and spin-down electrons have different energies, despite the absence of global magnetization. This unique combination arises from  specific crystal symmetry that allows alternating spin polarization while maintaining zero  total magnetic moment. Altermagnets are promising for spintronic applications because they offer fast magnetic switching and minimal stray fields, combining the advantages of ferromagnets (strong spin effects) with those of antiferromagnets (robustness against external fields). Numerous altermagnets have been theoretically predicted and experimentally verified\cite{k4,k5,k5-1,k6,k6-1,k6-2,k6-3,ex1,ex2,ex3,ex4}, which has significantly advanced the field of altermagnetism.

Hidden physics offers an alternative perspective for understanding material properties or overcoming certain forbidden electronic states\cite{h2,h3,h4,h5,h6,h7,h8}.
For instance, in certain materials, bulk Rashba spin splitting is absent; however, by introducing the degree of freedom of layers, such a phenomenon can emerge locally\cite{h3}. Furthermore, while half-metallicity is traditionally forbidden in conventional antiferromagnets (for example $PT$-antiferromagnets with the joint symmetry ($PT$) of space inversion symmetry ($P$) and time-reversal symmetry ($T$)) and altermagnets, the hidden half-metallicity may still exist\cite{h8}.
Recently, the hidden altermagnetism has been explicitly proposed: although no spin splitting is observed globally, the local altermagnetic (AM) spin splitting can still exist\cite{ha}. Subsequently, the multiferroic collinear antiferromagnet with hidden AM spin splitting is predicted in   $\mathrm{MnS_2}$, giving rise to various emergent responses\cite{ha1}. Tunable hidden AM spin splitting  has also been reported in layered Ruddlesden-Popper oxides\cite{ha2}.
The hidden altermagnetism has also been predicted to exist in $\mathrm{Sr_{n+1}Cr_n O_{3n+1}}$ due to orbital ordering rather than lattice symmetry\cite{ha2-1}.

\begin{figure}[t]
    \centering
    \includegraphics[width=0.45\textwidth]{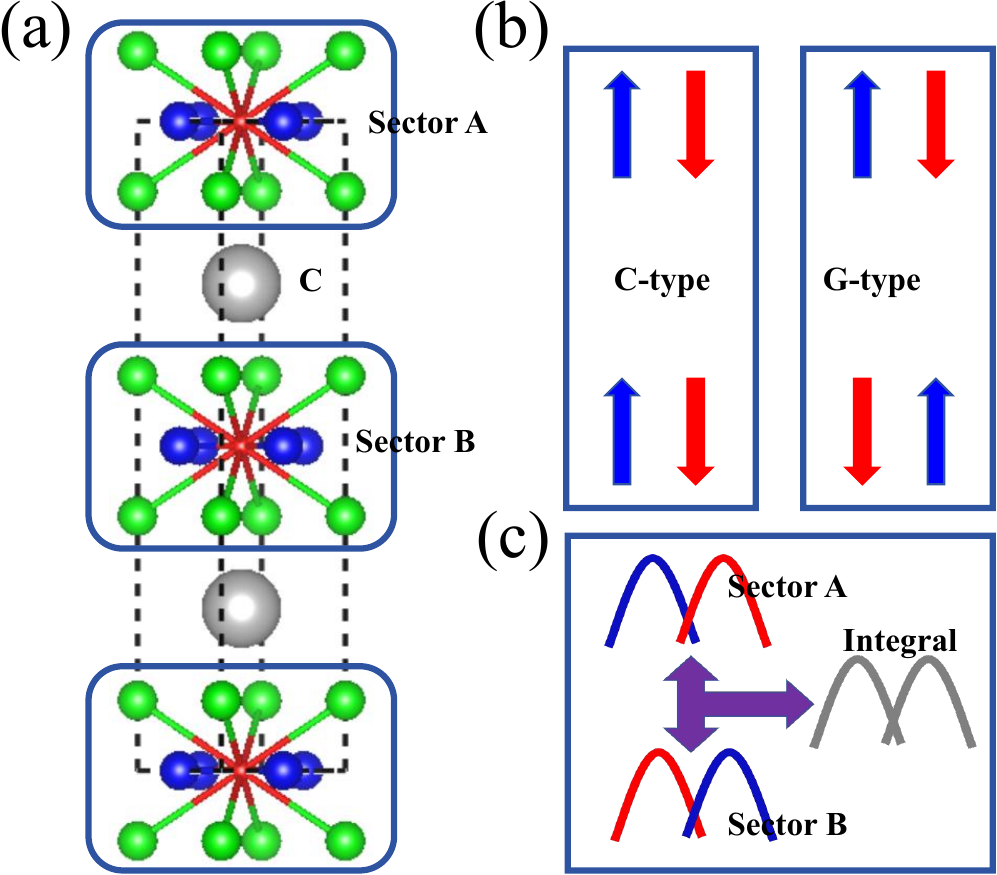}
    \caption{(Color online) For $\mathrm{CsV_2Te_2O}$, (a): the crystal structure with blue, red, green and gray spheres representing V, O, Te and Cs atoms, respectively. The black dashed box denotes the magnetic primitive cell, which consists of sector A, sector B, and C. (b): two
possible AFM  configurations  with  C-type  and G-type. (c): the  schematic diagram of hidden AM electronic state, and two sectors exhibit the
 altermagnetism with opposite spin polarization, yet the integral cancels
the spin polarization out.}\label{a}
\end{figure}

The hidden altermagnetism in $\mathrm{Cs_{1-\delta}V_2Te_2O}$ has recently been experimentally confirmed using neutron diffraction and spin-resolved angle-resolved photoemission spectroscopy (spin-ARPES)\cite{ha3}.  The Josephson effects in planar and vertical junctions based on $\mathrm{CsV_2Te_2O}$-family materials with hidden altermagnetism have been investigated, giving rise to  a general even-odd effect in hidden altermagnets\cite{ha4}. To simplify the calculation, we take
$\mathrm{\delta}$=0  in  $\mathrm{Cs_{1-\delta}V_2Te_2O}$, and  its crystal structure  is shown in \autoref{a} (a).  The two lowest-energy antiferromagnetic (AFM) configurations\cite{ha3,ha5}-the C-type (intralayer AFM, interlayer  ferromagnetic (FM)) and G-type (intralayer AFM, interlayer AFM)-are depicted in \autoref{a} (b), and the G-type AFM phase hosts a hidden AM electronic state (\autoref{a} (c)),  which has been experimentally confirmed.

Theoretically, these C-type and G-type  AFM configurations almost have the same energy with energy difference less than 0.5 meV per unit cell, and
 the G-type AFM ordering is the ground state only if the electronic correlation strength
$U$ is above a critical value\cite{ha3,ha5}. Nevertheless, the band structure at
$U$=0.00 eV  is closer to the experimental data\cite{ha3}.  Intrinsic Cs vacancies exist, and the energy difference of the two magnetic configurations  also depends on Cs vacancy distribution\cite{ha3,ha5}. 
Are there alternative methods to distinguish the C-type and G-type configurations, corresponding to apparent and hidden altermagnetism, besides directly identifying them via band structure measurements?
Here, we propose distinguishing these two magnetic configurations by applying in-plane uniaxial strain. For the C-type configuration, uniaxial strain induces a net magnetic moment, whereas the total magnetic moment of the G-type configuration always remains zero.
In fact, we  broaden the physical implications of the piezomagnetic effect\cite{k6}, which can be directly validated in experimentally realizable $\mathrm{KV_2Se_2O}$ and  $\mathrm{Rb_{1-\delta}V_2Te_2O}$\cite{ex3,ex4}.

\begin{figure*}[t]
    \centering
    \includegraphics[width=0.8\textwidth]{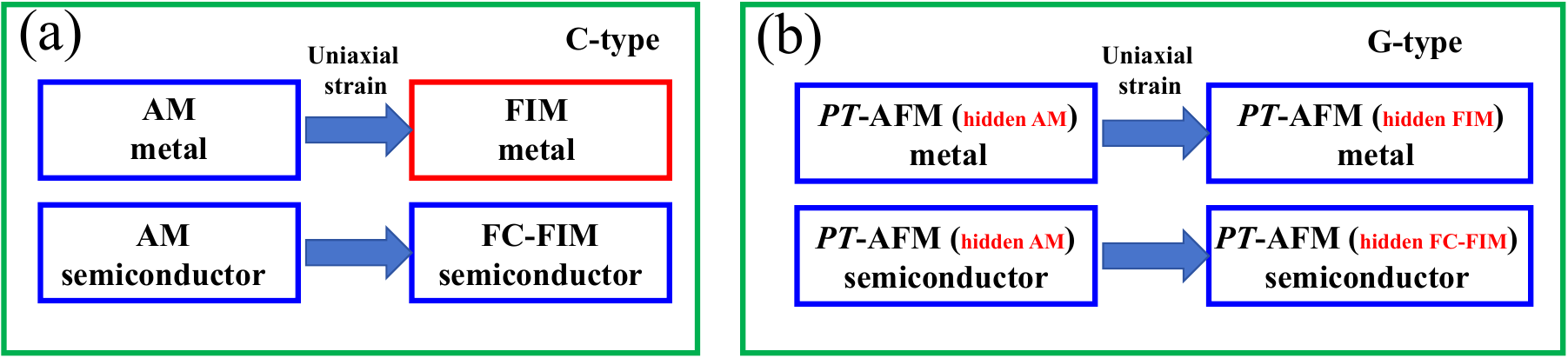}
    \caption{(Color online) For  $\mathrm{CsV_2Te_2O}$, under uniaxial strain, possible electronic state transitions with C-type (a) and G-type (b) AFM configurations. Blue boxes indicate zero total magnetic moment, while red box indicates  non-zero total magnetic moment. In (b), the red-colored text in parentheses represents the locally viewed electronic states. }\label{b}
\end{figure*}

\textcolor[rgb]{0.00,0.00,1.00}{\textbf{Fundamental strategy.---}}
The space group of magnetic  crystal structure of $\mathrm{CsV_2Te_2O}$  is  $P4/mmm$ (No.123), which possesses lattice $P$ symmetry\cite{ha3}. Then, such a lattice structure is  consistent with our originally proposed hidden AM model (see Fig.1 in ref.\cite{ha}). Therefore, it can be viewed as being composed of sector A, sector B, and  C (see \autoref{a} (a)). It is found that, by reversing the $\mathrm{N\acute{e}el}$ vector of sector B, the transition from C-type to G-type can be achieved  (see \autoref{a} (b)), that is, the transition from  apparent to hidden altermagnetism due to the increased symmetry operations complying with $PT$ symmetry. From an overall perspective, the hidden altermagnetism is essentially the conventional $PT$-antiferromagnetism (It should be noted that hidden  altermagnetism can be detected in $\mathrm{CsV_2Te_2O}$  via spin-ARPES because the system is highly quasi-two-dimensional, which allows the local spin polarization to be measured.).  When in-plane uniaxial strain is applied, for example along the $x$-direction, the space group of $\mathrm{CsV_2Te_2O}$  changes to $Pmmm$ (No.47), still preserving lattice  $P$ symmetry.  Below, we will  discuss the electronic state transitions upon applying uniaxial strain, categorized by C-type and G-type configurations as well as electronic structure characteristics (metal and  semiconductor).

For the C-type (see \autoref{b} (a)), under uniaxial strain, the $[C_2||C_4]$ (The $C_2$ denotes a twofold rotation in spin space, while the  $C_4$ represents a fourfold rotation in lattice space.) symmetry will be broken, and the  magnetic atoms with opposite spins lose their symmetric connection. If it is a metal, it will generally transform from an AM metal to a ferrimagnetic (FIM) metal; if it is a semiconductor, it will become a fully-compensated ferrimagnetic (FC-FIM) semiconductor, possessing  a zero magnetic moment guaranteed by proper electron filling rather than symmetry\cite{fc}. For the semiconductor case, this transition is strictly valid because the numbers of spin-up and spin-down electrons remain  unchanged, provided that the energy gap remains open; therefore, the total magnetic moment is strictly zero, that is, it becomes the so-called FC-FIM case. For the metal  case, under uniaxial strain, the numbers of spin-up and spin-down electrons will be adjusted. It is generally difficult to maintain the balanced state, which will lead to a net magnetic moment, resulting in the FIM  case. For a special type of semimetal in metals,  it may transform into an FC-FIM semimetal, and the numbers of spin-up and spin-down electrons  may  still be adjusted to a balanced state, continuing to maintain zero net magnetic moment, becoming the so-called FC-FIM case, which has already been confirmed in monolayer AM semimetal $\mathrm{Cr_2O}$\cite{fc1}.

For the G-type (see \autoref{b} (b)), under uniaxial strain, the spin-up and spin-down atoms can still be connected through $P$ symmetry, and the $PT$ symmetry is still preserved. Regardless of whether it is metal and  semiconductor, the electronic state remains  $PT$-AFM. Therefore, under uniaxial strain, the total magnetic moment remains zero in all cases. However, viewed locally at sectors A and B, the magnetic atoms with opposite spins have no symmetric connection.  From a local perspective, under uniaxial strain, the  electronic state transitions for metallic and semiconducting phases proceed as follows: the hidden AM metal transforms into a hidden FIM metal, while the hidden AM semiconductor evolves into a hidden FC-FIM semiconductor, respectively. The transition of the local  electronic state of individual sector A or B under uniaxial strain is similar to the discussion of the C-type case above.
\begin{figure}[t]
    \centering
    \includegraphics[width=0.45\textwidth]{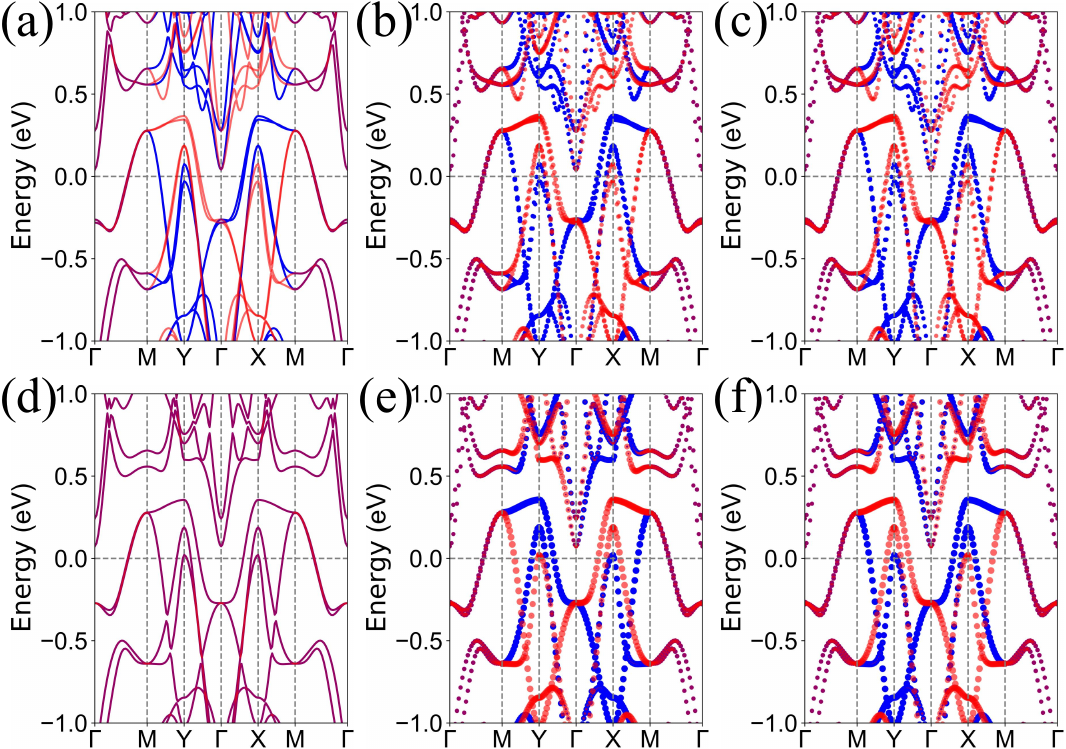}
     \caption{(Color online) For  $\mathrm{CsV_2Te_2O}$ without uniaxial strain,  the global  energy  band structure (a, d)  along with the spin-resolved projections onto the sector A (b, e) and sector B (c, f) with C-type (a, b, c) and G-type (d, e, f) AFM configurations. The blue, red, and purple curves denote the spin-up, spin-down, and spin-degenerate bands, respectively.  In (b, c, e, f), the weighting coefficient is proportional to the circle size.}\label{c}
\end{figure}

\begin{figure*}[t]
    \centering
    \includegraphics[width=0.90\textwidth]{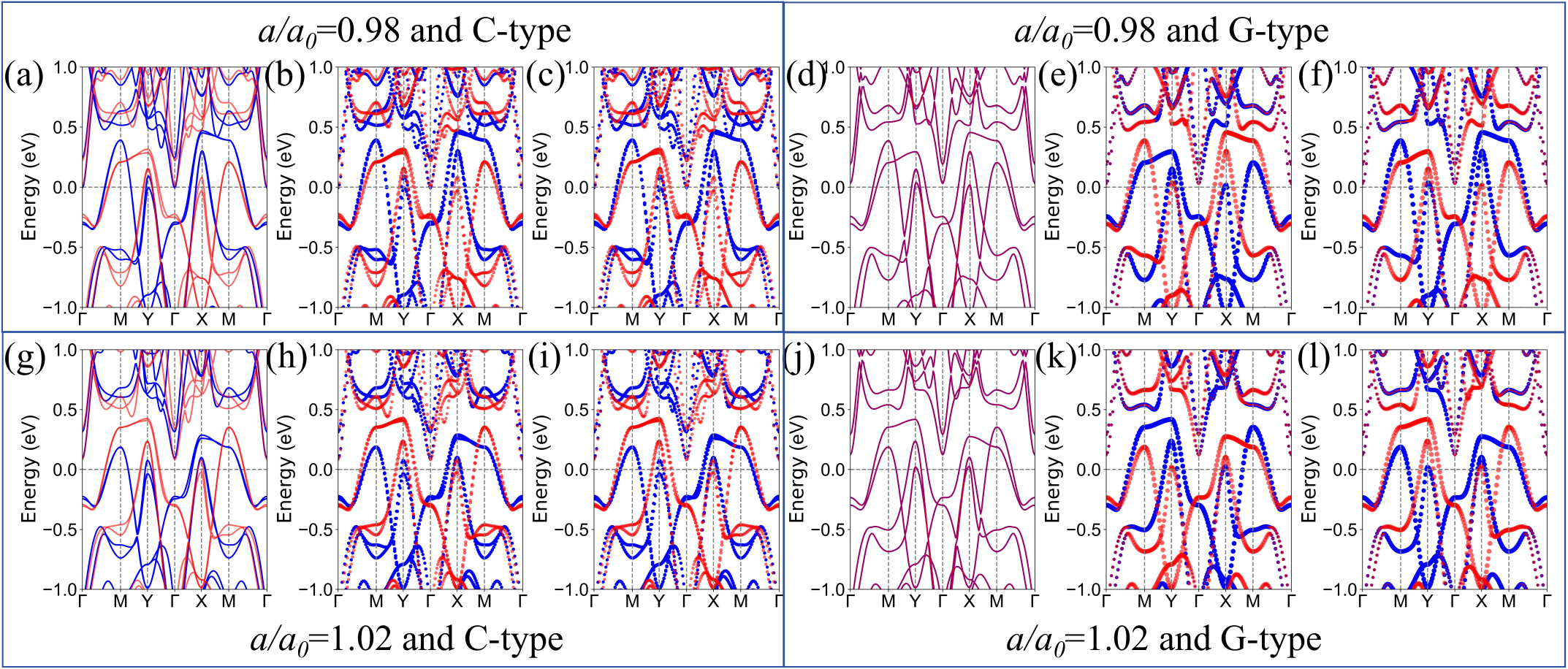}
     \caption{(Color online)  For  $\mathrm{CsV_2Te_2O}$ with $a/a_0$=0.98 (a, b, c, d, e, f) and 1.02 (g, h, i, j, k, l),  the global  energy  band structure (a, d, g, j)  along with the spin-resolved projections onto the sector A (b, e, h, k) and sector B (c, f, i, l) with C-type (a, b, c, g, h, i) and G-type (d, e, f, j, k, l) AFM configurations. The blue, red, and purple curves denote the spin-up, spin-down, and spin-degenerate bands, respectively.  In (b, c, e, f, h, i, k, l), the weighting coefficient is proportional to the circle size.}\label{d}
\end{figure*}
\begin{figure}[t]
    \centering
    \includegraphics[width=0.45\textwidth]{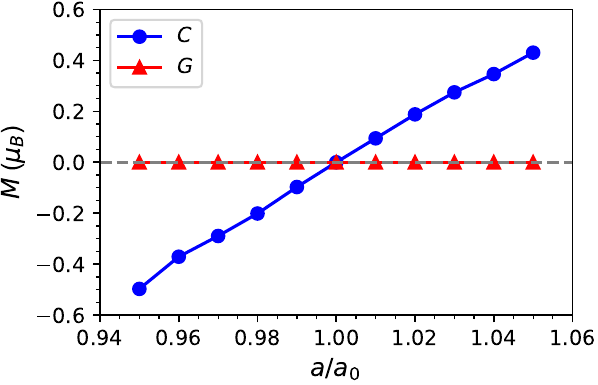}
     \caption{(Color online)  For  $\mathrm{CsV_2Te_2O}$, the total magnetic moment as a function of $a/a_0$ with C-type  and G-type AFM configurations.}\label{e}
\end{figure}
\begin{figure*}[t]
    \centering
    \includegraphics[width=0.90\textwidth]{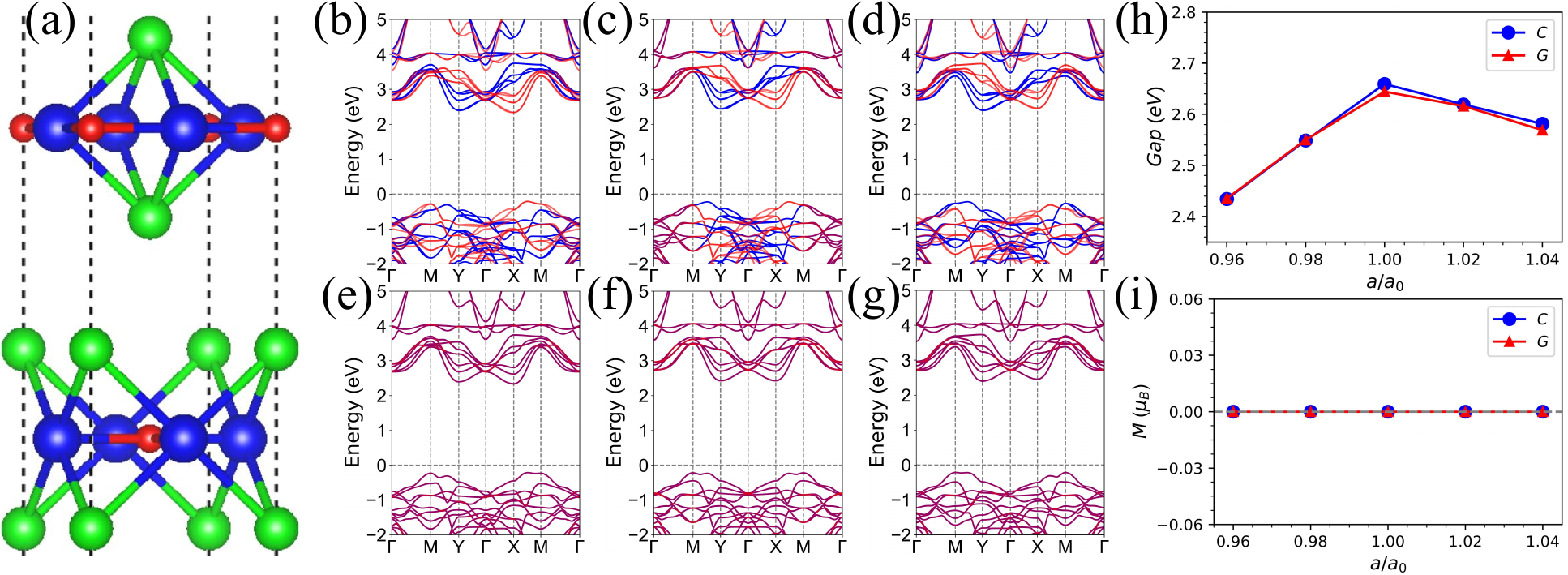}
     \caption{(Color online)  For  $\mathrm{Fe_2Br_2O}$ bilayer,  the crystal structure, where blue, green, and red spheres represent Fe, Br, and O atoms, respectively (a); the global energy band structures at $a/a_0$=0.98 (b, e), 1.00 (c, f),  and 1.02 (d, g) with C-type (b, c, d) and G-type (e, f, g) AFM configurations; the total band gap (h) and total magnetic moment (i) as functions of $a/a_0$ with  C-type and G-type configurations. In (b, c, d, e, f, g), the blue, red, and purple curves denote the spin-up, spin-down, and spin-degenerate bands, respectively.}\label{f}
\end{figure*}
\begin{figure}[t]
    \centering
    \includegraphics[width=0.45\textwidth]{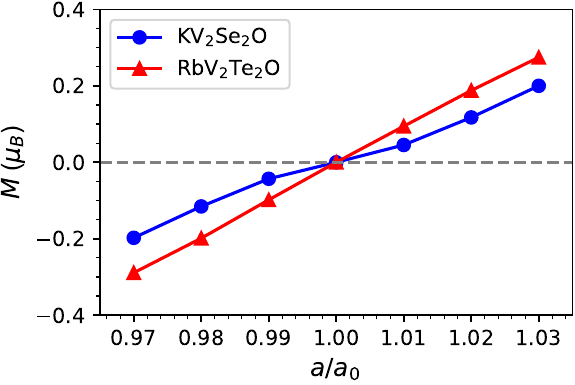}
     \caption{(Color online)  For  $\mathrm{KV_2Se_2O}$ and  $\mathrm{RbV_2Te_2O}$, the total magnetic moment as a function of $a/a_0$ with C-type  AFM configuration.}\label{g}
\end{figure}
According to \autoref{b}, a net magnetic moment can be induced by uniaxial strain only in the C-type configuration for metallic system. For  $\mathrm{CsV_2Te_2O}$, which exhibits pronounced metallic behavior, the application of uniaxial strain should generate  a net magnetic moment in the C-type configuration but not in the G-type configuration, thereby distinguishing apparent and hidden altermagnetism. The phenomenon of inducing a net magnetic moment in altermagnetism by uniaxial strain proposed here can be termed the piezomagnetic effect. However, this differs from the semiconductor cases studied in existing literature, where uniaxial strain must first induce a transition from the AM to FC-FIM electronic state, followed by carrier doping to generate a net magnetic moment\cite{k6}.

\textcolor[rgb]{0.00,0.00,1.00}{\textbf{Computational detail.---}}
Density functional theory (DFT) calculations\cite{1,111} are performed using the Vienna ab initio simulation package (VASP)\cite{pv1,pv2,pv3} within the framework of the projector augmented-wave (PAW) method. The generalized gradient approximation (GGA) proposed by Perdew, Burke, and Ernzerhof (PBE)\cite{pbe} is employed as the exchange-correlation functional.
A kinetic energy cutoff of 500 eV, a total energy convergence criterion of $10^{-8}$ eV, and a force convergence criterion of 0.001$\mathrm{eV\cdot{\AA}^{-1}}$ are adopted. A 13$\times$13$\times$2 Monkhorst-Pack $k$-point mesh is used to sample the Brillouin zone (BZ) for both structural relaxation and electronic structure calculations.  For $\mathrm{Fe_2Br_2O}$, the Hubbard correction is incorporated within the rotationally invariant approach proposed by Dudarev et al.\cite{du}. For the bilayer $\mathrm{Fe_2Br_2O}$, a vacuum slab thicker than 15 $\mathrm{{\AA}}$ is introduced to eliminate spurious interactions between periodic supercells, and the dispersion-corrected DFT-D3 method\cite{dft3} is utilized to account for van der Waals interactions. When uniaxial strain is applied along the $a$-axis, both the $b$ and $c$ axes are fully optimized for bulk materials. In contrast, only the $b$-axis can be optimized in the bilayer system.

\textcolor[rgb]{0.00,0.00,1.00}{\textbf{Material verification.---}}
In our originally proposed hidden AM candidate material, the structure is realized by directly stacking square AM monolayers of $\mathrm{Cr_2SO}$ in an AA configuration\cite{ha}. However, AA stacking is not the lowest-energy configuration. Similarly, the experimentally verified $\mathrm{CsV_2Te_2O}$ is  also constructed by directly AA-stacking square AM monolayers of $\mathrm{V_2Te_2O}$, with the intercalated Cs layers serving to stabilize the AA configuration\cite{ha3}.
This provides a fundamental framework for exploring the  potential hidden AM materials.
The optimized parameters are $a$=$b$=4.087 $\mathrm{{\AA}}$  and  $c$=17.822 $\mathrm{{\AA}}$ with the magnetic primitive cell, which  are in good agreement with existing experimental and theoretical results\cite{ha3,ha5}.

For  unstrained  $\mathrm{CsV_2Te_2O}$,  the global  energy  band structure   along with the spin-resolved projections onto the sector A and sector B with C-type  and G-type  AFM configurations are plotted in \autoref{c}. For the C-type, as the $[C_2||C_4]$ symmetry is satisfied, the band structure exhibits a $d$-wave AM metallic state. The total magnetic moment is strictly zero, as guaranteed by symmetry. The spin-resolved band structures projected onto sectors A and B indicate that these two sectors are completely equivalent. If each sector yields a net magnetic moment, the moments of two sectors should be equal in magnitude and add constructively.
For the G-type, due to $PT$ symmetry, the energy bands remain degenerate across the entire BZ. The symmetry also ensures a strictly vanishing total magnetic moment.  Nevertheless, neglecting the interaction between the sectors, the spin-resolved band structures projected onto sectors A and B show that the two sectors possess  $d$-wave AM metallic properties with opposite spin polarizations. If each sector generates a net magnetic moment, the moments of two sectors  should be equal in magnitude but cancel each other out.

The influences  of uniaxial strain along the $x$ direction on the electronic state of $\mathrm{CsV_2Te_2O}$ are  investigated  by employing
$a/a_0$ (0.95$\sim$1.05) to model the strain, where $a$ and $a_0$ represent the lattice parameters of the strained and equilibrium structures, respectively.
 For  $\mathrm{CsV_2Te_2O}$ with $a/a_0$=0.98  and 1.02,  the global  energy  band structure  along with the spin-resolved projections onto the sector A  and sector B  with C-type  and G-type  AFM configurations are plotted in \autoref{d}. Upon applying uniaxial strain, for the C-type case, the spin splitting exists throughout the entire BZ in both the total band structure and the band structures corresponding to the two sectors, exhibiting the so-called $s$-wave symmetry.
 This differs from unstrained $d$-wave altermagnetism, which is inherently spin-degenerate at the $\Gamma$ point\cite{k4}. After uniaxial strain is applied, combined with a nonzero total magnetic moment, the electronic state transitions to FIM case. For the G-type magnetic structure, the overall band structure is still  spin-degenerate. However, the spin splitting occurs in the two sectors and exhibits $s$-wave symmetry, corresponding to FIM case. Since their spin polarizations are opposite, the net magnetic moment vanishes. Upon applying uniaxial strain, the system remains a $PT$-antiferromagnet globally, but locally it exhibits the so-called hidden FIM electronic state. For all cases, uniaxial strain induces asymmetry in the band structures along the M-Y-$\Gamma$ and M-X-$\Gamma$ paths. This asymmetry is reversed as the strain changes from compressive to tensile.

As shown in \autoref{e}, under uniaxial strain, the total magnetic moment of the G-type magnetic structure always  remains zero, whereas for the C-type structure, the total magnetic moment increases almost linearly from negative to positive values as the strain changes from compressive to tensile.
Thus, by applying uniaxial strain to $\mathrm{CsV_2Te_2O}$ and measuring the magnetic moment, we can distinguish between C-type and G-type magnetic ordering, namely the distinct apparent and hidden altermagnetism. Here, the magnetic moment induced by uniaxial strain (i.e., the piezomagnetic effect) is one order of magnitude larger than that induced by first applying strain and then performing carrier doping in AM semiconductor\cite{k6}. When a strain of -5\% ($a/a_0$-1) is applied to $\mathrm{V_2Se_2O}$ and sufficient carriers are doped to saturation, the magnetic moment remains less than 0.03 $\mu_B$\cite{k6}.
In contrast, applying only -1\% ($a/a_0$-1) strain yields a total magnetic moment  of -0.10 $\mu_B$ in C-type $\mathrm{CsV_2Te_2O}$, which is more favorable for experimental measurement and verification.

\textcolor[rgb]{0.00,0.00,1.00}{\textbf{Discussion and Conclusion.---}}
For semiconductor systems, we also verify with concrete example that both C-type and G-type AFM configurations maintain a zero total magnetic moment under uniaxial strain. By adopting the originally proposed approach\cite{ha}, we construct crystal structures with hidden altermagnetism through the stacking of square AM  semiconductors.
Monolayer  $\mathrm{Fe_2Br_2O}$ has been verified to be a  stable square AM semiconductor (see FIG. S1\cite{bc})\cite{dc1}. Then, we construct an AA-stacked bilayer $\mathrm{Fe_2Br_2O}$, and by translating along vector ($\vec{a}$+$\vec{b}$)/2 , obtain the energetically favorable AC-stacked  configuration, whose structure is presented in \autoref{f} (a). The global energy band structures at $a/a_0$=0.98, 1.00  and 1.02  with C-type  and G-type  AFM configurations along with  the total band gap as a function of $a/a_0$ are plotted in  \autoref{f} (b-g and h).
Within the considered strain range, bilayer  $\mathrm{Fe_2Br_2O}$ remains semiconducting for all cases. For the C-type configuration, the application of uniaxial strain induces an electronic phase transition from the AM state to the FC-FIM state, accompanied by the emergence of spin-dependent valley polarization. Moreover, the valley polarization can be reversed upon switching the strain from compressive to tensile. For the G-type configuration, upon applying uniaxial strain, the system retains $PT$-antiferromagnetism globally, and  exhibits spin-independent valley polarization. In addition, switching the strain from compressive to tensile can also induce valley polarization reversal. Locally, it undergoes an electronic transition from hidden AM to hidden FC-FIM states. According to  \autoref{f} (i), for both C-type and G-type configurations, the total magnetic moment remains zero throughout the considered range of uniaxial strain, which is consistent with our proposal for semiconductor systems.

The experimentally synthesized and well-defined apparent  AM $\mathrm{KV_2Se_2O}$ and  $\mathrm{Rb_{1-\delta}V_2Te_2O}$ with C-type AFM configuration have been reported\cite{ex3,ex4}. Nevertheless, another experimental study using neutron diffraction techniques has revealed that $\mathrm{KV_2Se_2O}$  exhibits a G-type AFM configuration rather than a C-type one\cite{kkk}.  Since ARPES is a surface-sensitive technique, and measurements of altermagnetism by ARPES suffer from domain effects, it is not straightforward to distinguish between apparent and hidden altermagnetism based solely on ARPES spectra. Therefore, our proposed direct piezomagnetic effect can also  be used to verify apparent and hidden altermagnetism in $\mathrm{KV_2Se_2O}$ and  $\mathrm{Rb_{1-\delta}V_2Te_2O}$. The band structures of $\mathrm{KV_2Se_2O}$ and  $\mathrm{RbV_2Te_2O}$ with C-type AFM configuration under typical strains are presented in FIG.S2\cite{bc}, and the total magnetic moment as a function of strain is plotted in \autoref{g}. The corresponding calculated results are similar to those of the C-type AFM $\mathrm{CsV_2Te_2O}$, and  the uniaxial strain can also induce a net magnetic moment in both materials.  These are amenable to direct experimental validation.

In summary, we demonstrate that in-plane uniaxial strain enables an effective distinction between apparent and hidden altermagnetism in $\mathrm{CsV_2Te_2O}$. Under such strain, apparent altermagnetism produces a finite net magnetic moment, whereas hidden altermagnetism retains zero net magnetization. The strain-induced magnetization observed herein originates from the piezomagnetic effect, which differs fundamentally from the magnetism induced in semiconductor systems, where net magnetization requires sequential strain engineering and carrier doping. Our work enriches the fundamental understanding of strain-magnetism coupling mechanisms and facilitates the development of spintronic applications.

\begin{acknowledgments}
This work is supported by Natural Science Basis Research Plan in Shaanxi Province of China   (2025JC-YBMS-008). We are grateful to Shanxi Supercomputing Center of China, and the calculations were performed on TianHe-2.
\end{acknowledgments}

\end{document}